\documentclass[a4paper,eqsecnum,nofootinbib,showpacs]{revtex4}
\usepackage{slashed,amsmath,amssymb} 
\usepackage{epsfig,slashed}

\newcommand{\rar}{\rightarrow}
\newcommand{\lar}{\leftarrow}
\newcommand{\lra}{\leftrightarrow}

\newcommand{\nn}{\nonumber}
\newcommand{\beq} {\begin{equation}}
\newcommand{\eeq} {\end{equation}}
\newcommand{\beqa} {\begin{eqnarray}}
\newcommand{\eeqa} {\end{eqnarray}}

\newcommand{\ie}{{\it i.e.}}
\newcommand{\eg}{{\it e.g.}}

\newcommand{\rhs}{{\it rhs.}}

\newcommand{\eps}{\epsilon}
\newcommand{\ieps}{i\varepsilon}
\newcommand{\vphi}{\varphi}
\newcommand{\veps}{\varepsilon}
\newcommand{\order}[1]{${\cal O}\left(#1 \right)$}

\newcommand{\eq}[1]{(\ref{#1})}

\newcommand{\inv}[1]{\frac{1}{#1}}
\newcommand{\halft}{{\textstyle \frac{1}{2}}}

\newcommand{\ket}[1]{\left\vert{#1}\right\rangle}
\newcommand{\bra}[1]{\langle{#1}\vert}

\newcommand{\com}[2]{\left[{#1},{#2}\right]}
\newcommand{\acom}[2]{\left\{{#1},{#2}\right\}}

\newcommand{\bs}[1]{\boldsymbol{#1}}

\newcommand{\psl}{{\slashed{p}}}

\newcommand{\pv}{{\bs{p}}}

\newcommand{\bck}{\Phi}

\newcommand{\dms}{(m_1^2-m_2^2)}

\newcommand{\lrder}{{\buildrel\lra\over{\partial}}}

\begin{document}
{\par\raggedleft \texttt{CCTP-2012-01}\par}
{\par\raggedleft \texttt{HIP-2012-04/TH}\par}
\bigskip{}

\title{Boosting equal time bound states}

\author{Dennis D. Dietrich}
\affiliation{Institut f\"ur Theoretische Physik, Goethe-Universit\"at, Max-von-Laue-Str.~1,
D-60438 Frankfurt am Main, Germany}

\author{Paul Hoyer}
\affiliation{Department of Physics and Helsinki Institute of Physics\\ POB 64, FIN-00014 University of Helsinki, Finland}

\author{Matti J\"arvinen}
\affiliation{Crete Center for Theoretical Physics\\ 
Department of Physics, University of Crete\\71003 Heraklion, Greece}

\begin{abstract}
We present an explicit and exact boost of a relativistic bound state defined 
at equal time of the constituents 
in the Born approximation (lowest order in $\hbar$). 
To this end, we construct the Poincar\'e generators of QED and QCD in $D=1+1$ dimensions, using Gauss' law to express $A^0$ in terms of the fermion fields in $A^1=0$ gauge. We determine the fermion-antifermion bound states in the Born approximation 
as eigenstates of the time and space translation generators $P^0$ and $P^1$. The boost operator is combined with a gauge transformation so as to maintain the gauge condition $A^1=0$ in the new frame. We verify that the boosted state remains an eigenstate of $P^0$ and $P^1$ with appropriately transformed eigenvalues and determine the transformation law of the equal-time, relativistic wave function. The shape of the wave function is independent of the CM momentum when expressed in terms of a variable which is quadratically related to the distance $x$ between the fermions. As a consequence, the Lorentz contraction of the wave function is $\propto 1/(E-V(x))$ and thus depends on $x$ via the linear potential $V(x)$.

\end{abstract}


\maketitle

%%%%%%%%%%%%%%%%%%%%%%
\section{Introduction}
%%%%%%%%%%%%%%%%%%%%%%

Physical gauge field theories are Poincar\'e invariant, \ie, their action is symmetric under space-time translations, rotations and boosts. In a Hamiltonian treatment
the quantization surface is, however, not left invariant by all 10 Poincar\'e generators. An equal-time ($t=0$) surface is invariant under space translations and rotations, which allows to construct explicit and exact eigenstates of the 3-momentum $\bs{P}$ and angular momentum $\bs{J}$ operators: These symmetries are ``kinematic''. Time translations and boosts on the other hand transform the equal-time surface. Those symmetries are termed ``dynamic'' and in practice cannot be implemented exactly \cite{Dirac:1949cp}. Thus the eigenstates of the time translation operator $P^0 = H$ (the Hamiltonian) can usually be found only in some approximation. Similarly, the boost operators are dynamic operators, which create and destroy particles.

The gauge coupling $\alpha$ is a free parameter of the Lagrangian. This ensures the Poincar\'e invariance of each order of a perturbative expansion of the S matrix. In a time-ordered expansion boost invariance is obtained only after summing over all states at any intermediate time (at the given order of $\alpha$). Relativistic bound states have an infinite number of Fock components. Consequently their equal-time wave functions (which contain all powers of $\alpha$) transform in a highly non-trivial way under boosts. 
In practice,
no explicit, exact relation between the wave functions of relativistic states in different frames can be found: Boosting an equal-time state is as difficult as finding the eigenstates of the Hamiltonian directly in the new frame. 

At the lowest order in $\alpha$ scattering amplitudes are given by tree diagrams, whose internal propagators are off-shell. These Born amplitudes are independent of the $\ieps$ prescription used in the propagators. For retarded propagators $S_R(p^0,\pv)$, with poles at $p^0=\pm \sqrt{\pv^2+m^2}-\ieps$,  intermediate states of both positive and negative energy move forward in time: $S_R(t,\pv) \propto \theta(t)$. The absence of backward propagation (``$Z$'' diagrams) 
avoids intermediate pairs 
and thus simplifies time ordering, without changing the actual values of the scattering amplitudes in the Born approximation. 

Analogously, the bound state energies of an electron in a static external Coulomb potential are (at tree-level) independent of the $\ieps$ prescription used in the electron propagator.
In order to determine the equal-time Fock structure of the bound states,
\ie, the wave function of the electron, one needs to time-order the propagators. The time-ordering of Feynman propagators gives a wave function with any number of $e^+e^-$ pair components, which arise from $Z$ diagrams.
Using retarded propagators there are no $Z$-contributions and one obtains the standard Dirac wave function describing a single particle with both positive and negative energy components. Remarkably, the same relativistic bound state can thus, at Born level, be equivalently described using two quite different wave functions \cite{Hoyer:2009ep}. The $\ieps$-prescription invariance ensures that the bound state energies
are independent of the choice of wave function.

The possibility to describe relativistic bound states, which have an infinite sea of constituents, using few-particle ``valence'' wave functions reopens the issue of explicit Poincar\'e covariance. Since $\hbar$ is a fundamental parameter of the Lagrangian each order in an $\hbar$ expansion, and in particular the Born term, will have exact Poincar\'e invariance \cite{Brodsky:2010zk}. In the present paper we discuss cases where the wave functions of relativistic 
bound states in different frames can thus be related explicitly.

The Dirac bound states mentioned above are not translation invariant due to the external potential. We need to consider freely moving bound states formed by the interaction between two (or more) particles, such as QED atoms. The equal-time wave function of an atom in motion was considered in \cite{Jarvinen:2004pi}. In the rest frame, and at lowest order in $\alpha$ (and $\hbar$) the interaction is given by the standard Coulomb potential $V(r)=-\alpha/r$. Since atoms are non-relativistic the $Z$ diagrams are suppressed also for Feynman propagators. After a boost, however, the interaction (in Coulomb gauge) acquires also a propagating, transverse photon component. Thus, in a moving positronium atom $\ket{e^+e^-\gamma}$ Fock states must be included even at lowest order in $\alpha$. Adding relativistic corrections will increase the number of Fock states, further complicating the transformation of atomic wave functions under boosts. 

An explicit transformation law for relativistic states can be found for QED and QCD in $D=1+1$ dimensions. Since there are no transverse photons the interaction is fully given by the instantaneous (non-propagating) $A^0$ field in Coulomb gauge. We derive below the Poincar\'e algebra for QED starting from the non-local fermionic action, which is obtained by eliminating the $A^0$ field. This demonstrates that only interactions via a linear potential between the fermions, as stipulated by QED, lead to a Poincar\'e-covariant theory. The fact that the Lorentz boost operator must involve a gauge transformation in order to keep the Coulomb gauge condition satisfied in the boost is also illustrated.

We show how (Born level) two-body eigenstates of the translation generators $P^0$ and $P^1$ may be found in QED, making use of retarded propagation in analogy to the Dirac case mentioned above \cite{Hoyer:2009ep}. We then apply the boost generator to obtain the bound state in another frame. The boosted state remains an eigenstate of $P^0$ and $P^1$, with appropriately transformed eigenvalues. The rate of Lorentz contraction of the wave function turns out to depend on the linear potential $V(x)$ and thus on the distance $x$ between the constituents. The boost covariance of bound states of two fermions interacting via a linear potential has been noticed before as a property of their bound state equation \cite{Hoyer:1986ei}. Here we present the derivation from first principles, by identifying the system with a relativistic two-fermion bound state of 1+1 dimensional QED (or QCD) in the Born approximation.

The rest of the paper is organized as follows. In Sect.~\ref{secalgebra} we discuss the Poincar\'e algebra for 1+1 dimensional QED in Coulomb gauge after integrating out the gauge bosons, in Sect.~\ref{secqed} we present the corresponding bound state equation, in Sect.~\ref{boostcov} we analyze the behavior of the bound-state wave function under boosts, and in Sect.~\ref{discussion} we conclude the paper. Appendix \ref{generators} contains details on the derivation of the generators of the Poincar\'e algebra from Sect.~\ref{secalgebra}, and in App.~\ref{qcdapp} we present the generalization to QCD, \ie, to non-Abelian gauge groups.

%%%%%%%%%%%%%%%%%%%%%%%%%%%%%%%%%%%%%%%%%%
\section{Poincar\'e generators of QED$_2$} \label{secalgebra}
%%%%%%%%%%%%%%%%%%%%%%%%%%%%%%%%%%%%%%%%%%

We shall work in Coulomb gauge (here equivalent to $A^1=0$) in order to avoid Fock states with
longitudinal photons. The QED action in $D=1+1$ for fermions of flavor $f$ is then
\beq\label{qedact0}
S=\int d^2x \Big[-\inv{2}\big(\partial_1 A^0\big)\big(\partial^1 A^0\big)+\sum_f\psi_f^\dag(x)\gamma^0\big(i\slashed{\partial}-m_f -e\gamma^0 A^0\big)\psi_f(x) \Big] .
\eeq
The equation of motion for $A^0$ (Gauss' law),
\beq
-\partial_1^2 A^0(x) =e \sum_f\psi_f^\dag\psi_f(x) ,
\eeq
allows to express $A^0$ in terms of the fermion fields,
\beq\label{a0expr}
A^0(x) = -\frac{e}{2}\sum_f \int dy^1 |x^1-y^1| \psi_f^\dag\psi_f(x^0,y^1),
\eeq
in the absence of a background field \cite{Coleman:1976uz}. Using this in the action \eq{qedact0} gives the non-local expression
\beq\label{qedact1}
S \equiv S_F+S_V=\sum_f\int d^2x\, \psi_f^\dag(x)\gamma^0\big(i\slashed{\partial}-m_f\big)\psi_f(x) +
\frac{e^2}{4}\sum_{f,f'}\int d^2x\, d^2y\, \delta(x^0-y^0) \psi_f^\dag\psi_f(x) |x^1-y^1| \psi_{f'}^\dag\psi_{f'}(y) .
\eeq
Since no approximations have been made this action must be invariant under time and space translations as well as boosts, generated by the operators $P^0, P^1$ and $M^{01}$, respectively. Let us review the derivation of the Poincar\'e generators for the non-local action \eq{qedact1}, adapting the standard procedures (see, \eg, Sect.~7.3 in \cite{406190}).

Consider the infinitesimal space translation
\beq \label{dpsispace}
 \psi_f(x^0,x^1) \to \psi_f\big(x^0,x^1-\epsilon(x^0) d\ell\big) \ ,
\eeq 
where the {\em a priori} arbitrary function $\epsilon(x^0)$ will be a constant for a true translation. 
The variation of the free fermion action is
\beq
 \delta S_F = -d\ell \sum_{f}\int d^2x\Big[ \psi_f^\dag(x)\gamma^0\big(i\slashed{\partial}-m_f\big)\epsilon(x^0) \partial_1 \psi_f(x) + \epsilon(x^0) \big(\partial_1 \psi_f^\dag(x)\big)\gamma^0\big(i\slashed{\partial}-m_f\big)\Big] \psi_f(x) ,
\eeq
where $\slashed{\partial}$ in the first term operates both on $\epsilon(x^0)$ and the fermion field. As seen by integrating the last term partially over $x^1$, all terms except the one where  $\slashed{\partial}$ differentiates $\epsilon(x^0)$ cancel. Thus
\beq
 \delta S_F = -i d\ell \sum_{f}\int d^2x\ \epsilon'(x^0) \psi_f^\dag(x) \partial_1 \psi_f(x) \ .
\eeq
The transformation of the potential term $S_V$ can be canceled by a shift of integration variables $x^1 \to x^1 + \epsilon(x^0) d\ell$ and $y^1 \to y^1 + \epsilon(x^0) d\ell$, since the $\delta$-function sets $x^0 =y^0$  and the potential $|x^1-y^1|$ as well as $\epsilon(x^0)$ are unchanged by the shift. Therefore $\delta S_V = 0$ and the variation of the action becomes
\beq \label{dSspace}
 \delta S =  d\ell \int dx^0\ \epsilon'(x^0) P^1 ,
\eeq
where we identified 
the generator for spatial translations
\beq\label{spacegen}
 P^1 = -i  \sum_{f}\int dx^1\ \psi_f^\dag(x) \partial_1 \psi_f(x) \ .
\eeq
Setting $\epsilon(x^0)\equiv1$ the generic transformation~\eq{dpsispace} becomes a standard space translation and the variation~\eq{dSspace} vanishes, which proves the covariance of the QED action under space translations. 

Let us then assume that the fermion fields satisfy their equation of motion. Since the variation of the action vanishes under every infinitesimal transformation of $\psi(x)$, the variation \eq{dSspace} now vanishes for any function $\epsilon(x^0)$. Therefore,
\beq
  0 =  d\ell \int dx^0\ \epsilon'(x^0) P^1 = -d\ell \int dx^0\ \epsilon(x^0) \frac{d}{dx^0} P^1 \ .
\eeq
Since $\epsilon(x^0)$ was arbitrary we conclude that $P^1$ is conserved,
\beq
 \frac{d}{dx^0} P^1 = 0 ,
\eeq
when $\psi(x)$ satisfies its equation of motion.

The analogous derivations of the time translation $P^0$ and boost generator $M^{01}$ are given in Appendix \ref{generators}. Since the gauge constraint $A^1=0$ is not invariant under boosts $M^{01}$ is actually a combination of a Lorentz boost and a gauge transformation\footnote{In any event, for massless gauge fields like the photon, Lorentz and gauge transformations are entwined at the fundamental level, as these fields only represent a faithful vector representation of the Lorentz group up to a gauge transformation. (See, \eg, Sect.~5.9 in \cite{406190}.)}. Denoting $P^0 = P^0_F+P^0_V$ and $M^{01} = M^{01}_F+ M^{01}_V$ the result is
\beqa
P^0_F &=& \sum_{f}\int dx^1\ \psi_f^\dag(x)(-i\gamma^0\gamma^1\partial_1+m_f\gamma^0)\psi_f(x) , \nn
\\[-2mm] \label{timegenqed} \\[-2mm]
P^0_V &=& -\frac{e^2}{4}\sum_{f,f'}\int dx^1 dy^1\  \psi_f^\dag\psi_f(x^0,x^1) |x^1-y^1| \psi_{f'}^\dag\psi_{f'}(x^0,y^1) , \nn \\[2mm]
M^{01}_F &=& x^0 P^1 +  \sum_{f}\int dx^1\ \psi_f^\dag(x)\Big[ x^1( i \gamma^0\gamma^1 \partial_1 - \gamma^0 m_f) + \frac{i}{2} \gamma^0\gamma^1 \Big] \psi_f(x) , \nn
\\[-2mm]\label{boostgenqed} \\[-2mm]
M^{01}_V &=& \frac{e^2}{8}\sum_{f,f'}\int dx^1 dy^1\  \psi_f^\dag\psi_f(x^0,x^1) (x^1+y^1)|x^1-y^1| \psi_{f'}^\dag\psi_{f'}(x^0,y^1) \nn\ .
\eeqa

Let us then check that these generators satisfy the $D=1+1$ Poincar\'e Lie algebra.
Using the anticommutation relation
\beq\label{canacom}
\acom{\psi_{f\alpha}(x^0,x^1)}{\psi_{f'\beta}^\dag(x^0,y^1)}=\delta{(x^1-y^1)}\delta_{ff'}\delta_{\alpha\beta}
\eeq
it is straightforward to verify that the free generators $P^0_F,\ P^1$, and $M^{01}_F$ indeed satisfy 
\beqa\label{plie}
 \com{P^0}{P^1}= 0 , \label{plie01}\  \hspace{2cm}
 \com{P^0}{M^{01}} = i P^1 , \  \hspace{2cm}
 \com{P^1}{M^{01}} = i P^0 
\eeqa
among themselves. It is also easy to see that $\com{P^0}{P^1}=0$ holds when the interactions are included since $P^0_V$ in \eq{timegenqed} is invariant under space translations.

Of the three contributions to $\com{P^0}{M^{01}}$ that involve interactions the term $\com{P^0_V}{M_V^{01}}=0$ since $P^0_V$ and $M_V^{01}$ involve neither derivatives nor Dirac matrices. The two other terms
\beqa\label{intlie}
\com{P^0_V}{M_F^{01}}&=& \frac{e^2}{2}\sum_{f,f'}\int dx^1dy^1 \big[\psi_f^\dag(x)i\gamma^0\gamma^1\psi_f(x)\big]\psi_{f'}^\dag\psi_{f'}(y)\,x^1 \frac{\partial}{\partial x^1}|x^1-y^1| , \nn\\
\com{P^0_F}{M_V^{01}}&=& -\frac{e^2}{4}\sum_{f,f'}\int dx^1dy^1 \big[\psi_f^\dag(x)i\gamma^0\gamma^1\psi_f(x)\big]\psi_{f'}^\dag\psi_{f'}(y)\, \frac{\partial}{\partial x^1}\big[(x^1+y^1)|x^1-y^1|\big]
\eeqa
cancel, $\com{P^0_V}{M_F^{01}}+\com{P^0_F}{M_V^{01}}=0$, which ensures $\com{P^0}{M^{01}} = i P^1$. In the third Lie algebra relation
\beq\label{intlie2}
\com{P^1}{M_V^{01}} = -\frac{ie^2}{8}\sum_{f,f'}\int dx^1dy^1 \big[\psi_f^\dag(x)\psi_f(x)\big]\big[\psi_{f'}^\dag\psi_{f'}(y)\big]\,\big[ \big(\frac{\partial}{\partial x^1} +\frac{\partial}{\partial y^1}\big)(x^1+y^1)|x^1-y^1|\big] =iP_V^0
\eeq
ensures $ \com{P^1}{M^{01}} = i P^0$ with interacting generators. Note that the generator algebra \eq{plie} is satisfied only for the linear potential specified by QED$_2$.

%%%%%%%%%%%%%%%%%%%%%%%%%%%%%%%%%%%%%%%%%%
\section{Two-body bound states in QED$_2$} \label{secqed}
%%%%%%%%%%%%%%%%%%%%%%%%%%%%%%%%%%%%%%%%%%

As mentioned in the Introduction and further explained in \cite{Hoyer:2009ep}, a relativistic gauge theory bound state may be described by valence-like Dirac-type wave functions provided instantaneous Coulomb exchange dominates and one uses retarded propagators. At the Born level (lowest order in $\hbar$) the bound state energies will agree with the result using Feynman propagators, even though the wave functions obtained with the two types of propagator are very different.

In QED$_2$ Coulomb interaction is ensured by the gauge condition $A^1=0$. Retarded propagation for fermions is achieved using the ``retarded'' vacuum (which is equivalent to removing the Dirac sea)
\begin{equation}
\ket{0}_R=N^{-1}\prod_{p^1}d^\dagger(p^1)\ket{0} ,
\end{equation}
where the product is over antifermion creation operators of all momenta $p^1$ and $N$ is a normalization factor. The Pauli exclusion principle implies
\beq\label{psiann}
\psi(x)\ket{0}_R = 0
\eeq
for all $x$. This ensures retarded propagation, $_R\bra{0}T\big[\psi(x)\bar\psi(0)\big]\ket{0}_R \propto \theta(x^0)$, and 
forbids intermediate pairs.
The unusual ``vacuum'' $\ket{0}_R$ should be understood as a method of selecting terms that contribute at lowest order in $\hbar$. For perturbative loop corrections the boundary condition needs to be adjusted correspondingly 
to allow single or multiple pair production.

We define our fermion-antifermion bound states of energy $E$ and momentum $k$ by\footnote{The present definition of the wave function is related to the $\chi(x)$ used in \cite{Hoyer:2009ep,Hoyer:1986ei} as $\bck(x)=\gamma^0 \chi(x)\exp[-i\vphi(x)]$.}
\beq \label{ketk}
 \ket{E,k}\equiv \int dx_1 dx_2\, \exp\big[\halft
ik(x_1+x_2)\big]\bar\psi_1(0,x_1)e^{i\varphi}\bck(x_1-x_2)\psi_2(0,x_2)\ket{0}_R \
.
\eeq
Since we are working at Born level we may assume the fermion flavors $f=1,2$ to be distinct. The boundary condition corresponding to \eq{psiann} in the case of two flavors is taken to be 
\beq\label{2fbound}
\psi_1(x)\ket{0}_R = \psi_2^\dag(x)\ket{0}_R = 0 .
\eeq 
In \eq{ketk} the space coordinate of fermion $j$ is denoted $x_j \equiv x_j^1$ ($j=1,2$), and the state is defined at equal time, $x_j^0 = 0$. The wave function is the product of a plane wave in the CM position coordinate $\halft(x_1+x_2)$ and a $2\times 2$ matrix function $\bck(x)e^{i\vphi(x)}$ of the relative coordinate $x\equiv x_1-x_2$. As we shall see below, the extraction of the phase $\vphi(x)$ makes
the transformation of the wave function $\bck(x)$ under boosts, \ie, its $k$ dependence, more easily tractable. 
The phase is defined by
\beq\label{phidef}
\vphi(x) = -\dms(\xi+\zeta)\frac{\veps(x)}{e^2} ,
\eeq
where 
$\veps(x) \equiv x/|x|$ is the 
sign
function. The standard boost parameter of the bound state (of rest mass $M$) is denoted $\xi$,
\beq\label{xidef}
\sinh\xi = \frac{k}{M} , \hspace{2cm} \cosh\xi = \frac{E}{M} ,
\eeq 
whereas $\zeta$ is defined by
\beq\label{zdef}
 \hspace{1cm} \sinh\zeta = -\frac{k}{\sqrt{p^2}} , \hspace{1.5cm} \cosh\zeta = \frac{E-V(x)}{\sqrt{p^2}} ,
\eeq 
and depends on the relative coordinate $x$ through the linear QED$_2$ potential
\beq\label{potdef}
 V(x) = \halft\,e^2|x| .
\eeq
The $x$-dependent ``momentum'' $p$ appearing in \eq{zdef},
\beq\label{pdef}
p \equiv (E-V,-k) , \hspace{2cm} \psl = (E-V)\gamma^0+k\gamma^1 , \hspace{2cm} p^2 = (E-V)^2-k^2 ,
\eeq
is obtained by a $\zeta$-boost from the rest frame,
\beq\label{pboost}
\psl = \exp(\halft\zeta\gamma^0\gamma^1)\sqrt{p^2}\gamma^0\exp(-\halft\zeta\gamma^0\gamma^1) .
\eeq
Equation \eq{phidef} extends the definition of the phase $\vphi(x)$ first found in \cite{Hoyer:1986ei} to $x<0$. We also add the $x$-independent term $\propto \xi$ which is required by the boost transformation to be studied in the next section. Since $V(0)=0$, $p^2 = E^2-k^2 = M^2$ and $\xi+\zeta=0$ at $x=0$. The parameter $\zeta$ in \eq{zdef}, and consequently the phase $\vphi(x)$, are, however, well-defined only for $p^2>0$. Therefore, we shall here restrict to the region near the origin, with $|x|<2(E-|k|)/e^2$, 
where $p^2$ remains positive. This range covers the whole wave function in the weak coupling limit $e \to 0$. Notice also that this is only a restriction of the ``covariant'' formulation involving the variable $\zeta$ and the particular choice of the phase $\vphi(x)$ in \eq{phidef}, whereas the Poincar\'e algebra defines the bound state equation and the transformation of the wave function for all values of the coordinates.

In $D=1+1$ dimensions we may represent the Dirac matrices in terms of the Pauli matrices as
\beq
\gamma^0=\sigma_3 , \hspace{2cm} \gamma^1=i\sigma_2 ,  \hspace{2cm} \gamma^0\gamma^1=\sigma_1 .
\eeq

Applying the space translation generator \eq{spacegen} we may verify that the state \eq{ketk} has total momentum $k$,
\beq
 P^1 \ket{E,k} = k \ket{E,k} .
\eeq
Using \eq{timegenqed} the energy eigenvalue condition $P^0\ket{E,k}=E\ket{E,k}$ gives a bound state equation for the wave function $\bck(x)$,
\beq \label{bse}
i\partial_x\acom{\sigma_1}{\bck(x)}-(\partial_x\vphi)\acom{\sigma_1}{\bck(x)}-\halft k\com{\sigma_1}{\bck(x)}+m_1\sigma_3\bck(x)-m_2\bck(x)\sigma_3 = \big[E-V(x)\big] \bck(x) , 
\eeq
where the $x$-derivative of $\vphi(x)$ at constant $k$ is given by \eq{phidef} as
\beq\label{phiderx}
\partial_x\vphi(x) \equiv \left.\frac{\partial\vphi}{\partial x}\right|_k = \dms\frac{k}{2p^2} .
\eeq
In terms of $p$, the bound state equation can be written as
\beq\label{bsep}
i\partial_x\acom{\sigma_1}{\bck(x)}-(\partial_x\vphi)\acom{\sigma_1}{\bck(x)}-\sigma_3\left(\halft \psl-m_1\right)\bck(x) -\bck(x)\left(\halft \psl+m_2\right)\sigma_3 = 0 .
\eeq
We wish to ascertain that the bound state energy has the correct $k$ dependence, $E=\sqrt{k^2+M^2}$. There is no previous experience (except \cite{Hoyer:1986ei}) of how the wave function $\bck$ should depend on $k$.

Since $\bck$ is a $2\times 2$ matrix it has four independent components, which may be taken to be the coefficients of the unit and Pauli matrices,
\beqa
\bck(x) &\equiv& \bck_0(x) + \sum_{j=1}^3 \bck_j(x)\sigma_j
        = \phi(x)+\bck_2(x)\sigma_2+\bck_3(x)\sigma_3 , \\[2mm]
\phi(x) &\equiv& \bck_0(x)+\bck_1(x)\sigma_1 . \label{phiwf}
\eeqa
As $\bck_2$ and $\bck_3$ do not contribute to the derivative $i\partial_x\acom{\sigma_1}{\bck}$ in the bound state equation, these two components can be expressed in terms of $\phi$. We find
\beq\label{chidef}
\bck(x) = \frac{\slashed{p}}{p^2}\big(\halft\slashed{p}+m_1\big)\phi
 + \phi\big(\halft\slashed{p}-m_2\big)\frac{\slashed{p}}{p^2}
= \phi + \frac{1}{p^2}
(m_1 \slashed{p}\, \phi-m_2\phi\, \slashed{p}) .
\eeq

The bound state equation can be expressed in a frame-independent way by introducing the new variable
\beqa\label{sdef}
s(x) &\equiv& \inv{2}\int_0^x du \big[E-V(u)\big] = \frac{\veps(x)}{2e^2}\big[2EV(x)-V(x)^2\big] = \frac{\veps(x)}{2e^2}(M^2-p^2) ,\\[2mm]
\frac{d s}{d x} &=& \left.\frac{\partial s}{\partial x}\right|_k = \inv{2}\big[E-V(x)\big] , \nn
\eeqa
where $M \equiv \sqrt{E^2-k^2}$ is the rest mass of the bound state. Then \eq{bsep} implies\footnote{For conciseness of notation we denote by $\Phi(s)$ the wave function $\Phi(x(s))$ implicitly defined by \eq{sdef}.}
\beq\label{bsephi}
i\partial_s\sigma_1\phi(s) = \left[1-\frac{m_1^2+m_2^2}{p^2}\right]\phi(s)+\frac{2m_1 m_2}{p^2} \sigma_3\phi(s)\sigma_3 .
\eeq
Equivalently, the bound state condition for the components $\Phi_0$ and $\Phi_1$ of the wave function are
\beq\label{psibses}
i\partial_s\bck_1(s) = \left[1-\frac{(m_1-m_2)^2}{p^2}\right]\bck_0(s) , \hspace{2cm} 
i\partial_s\bck_0(s) = \left[1-\frac{(m_1+m_2)^2}{p^2}\right]\bck_1(s) .
\eeq
The conditions \eq{bsephi} and \eq{psibses} are independent of the CM momentum $k$ since according to \eq{sdef} $p^2$ is $k$ independent at fixed $s$ and rest mass $M$. This means that $\Phi_0(s)$ and $\Phi_1(s)$, and hence also $\phi(s)$, are the same functions of $s$ in all reference frames. According to \eq{pboost} and \eq{chidef} the full wave function $\bck(s)$ of a bound state with momentum $k$ is given by the rest frame $(k=0)$ wave function $\bck^{(k=0)}(s)$ as
\beq\label{krel}
\bck(s) = e^{\sigma_1\zeta/2}\bck^{(k=0)}(s) e^{-\sigma_1\zeta/2} ,
\eeq
possibly up to an $s$-independent factor. The boost parameter $\zeta$ is given in \eq{zdef}. The relation between $s$ and $x$ is $k$ dependent and thus different for $\bck(s)$ and $\bck^{(k=0)}(s)$.

As seen from \eq{sdef} the bound state equation \eq{bse} gives the correct dependence between energy and momentum, $E=\sqrt{k^2+M^2}$, only for the linear potential of QED$_2$. This is ensured by the Lorentz invariance of the QED action and the expansion in $\hbar$. The frame independence of the wave function, when expressed as a function of $s$, was first observed in \cite{Hoyer:1986ei}. It implies that the Lorentz contraction of the bound state is $x$ dependent: $dx/ds=2/(E-V(x))$.
Non-relativistic wave functions ($V \ll E$) transform globally, with a $1/E$  
contraction \cite{Jarvinen:2004jf}.

%%%%%%%%%%%%%%%%%%%%%%%%%%%%%%%%%%%%%%%%%%%%%%%%%%%%%%%%%%%%%%%%
\section{Boost covariance of the wave function} \label{boostcov}
%%%%%%%%%%%%%%%%%%%%%%%%%%%%%%%%%%%%%%%%%%%%%%%%%%%%%%%%%%%%%%%%

In the previous section we found that the dependence on the CM momentum $k$ of the solutions to the bound state equation \eq{bse} are related as in \eq{krel} (up to an $x$-independent factor). We shall now demonstrate that this relation is consistent with a direct boost of the bound states, using the generator $M^{01}$ derived in Sect. \ref{secalgebra}.

The sign convention of the Lie algebra \eq{plie} implies that the state $\ket{E+d\xi k, k+d\xi E}$ of 2-momentum $\Lambda k$, corresponding to the infinitesimal boost defined by \eq{coordboost}, is generated by $-id\xi M^{01}$,
\beq\label{boostop}
P^\mu(1-id\xi M^{01}) \ket{E,k} = k^\mu(1-id\xi M^{01}) \ket{E,k} +id\xi \com{M^{01}}{P^\mu}\ket{E,k}
= (\Lambda k)^\mu(1-id\xi M^{01}) \ket{E,k} .
\eeq
From its definition \eq{ketk} the $k$ dependence of the wave function $\bck(x)$ at constant $x\equiv x_1-x_2$ is thus given by the boost operator through
\beqa\label{ketdk}
(1-id\xi M^{01}) \ket{E,k}&=&\ket{E+d\xi k, k+d\xi E}= \int dx_1 dx_2 e^{ik(x_1+x_2)/2+i\vphi(x)}  \nn\\
&\times& \bar\psi_1(0,x_1)\Big\{\bck+id\xi E\Big[\halft(x_1+x_2)+\left.\frac{\partial \varphi(x)}{\partial k}\right|_x\Big]\bck+d\xi E \left.\frac{\partial \bck(x)}{\partial k}\right|_x\Big\}\psi_2(0,x_2)\ket{0}_R m,
\eeqa
where the $k$ dependence of $\vphi(x)$ at constant $x$ is obtained from \eq{phidef},
\beq\label{phiderk}
\left.\frac{\partial \varphi(x)}{\partial k}\right|_x = \dms\,\frac{E-V}{2Ep^2}\,x .
\eeq

Using the representation \eq{boostgenqed} of $M^{01}$ we find (with $\partial_j \equiv \partial/\partial x_j$)
\beqa
M^{01}\ket{E,k} &=&\int dx_1 dx_2\bar\psi_1(0,x_1)\Big\{-\big[x_1\sigma_1 i{\buildrel\rar\over{\partial_1}}+x_1\sigma_3 m_1 +\halft i\sigma_1\big]e^{ik(x_1+x_2)/2+i\vphi(x)}\bck(x) \nn\\
&+& e^{ik(x_1+x_2)/2+i\vphi(x)} \bck(x)\big[i{\buildrel\lar\over{\partial_2}}x_2\sigma_1+x_2\sigma_3 m_2 +\halft i\sigma_1-\halft(x_1+x_2)V(x)\big]\Big\}\psi_2(0,x_2)\ket{0}_R \nn\\[2mm]
&=& \int dx_1 dx_2 e^{ik(x_1+x_2)/2+i\vphi(x)}\bar\psi_1(0,x_1) \\
&\times&\Big\{\halft(x_1+x_2) \big(-i\partial_x\acom{\sigma_1}{\bck}+(\partial_x\vphi)\acom{\sigma_1}{\bck}+\halft k \com{\sigma_1}{\bck}-m_1\sigma_3 \bck +m_2 \bck\sigma_3-V\bck\big) \nn\\
&+& \halft x \big(-i\partial_x\com{\sigma_1}{\bck}+(\partial_x\vphi)\com{\sigma_1}{\bck}+\halft k \acom{\sigma_1}{\bck}-m_1\sigma_3 \bck -m_2 \bck\sigma_3\big)-\halft i\com{\sigma_1}{\bck}\Big\}\psi_2(0,x_2)\ket{0}_R .\nn
\eeqa
The coefficient of $\halft(x_1+x_2)$ on the next-to-last line equals $-E\bck$ by the bound state equation \eq{bse}. Hence it cancels against the corresponding term on the \rhs\ of \eq{ketdk}. The remaining terms specify the $k$ dependence of the wave function (at fixed $x$) implied by the boost operator,
\beqa\label{chider1}
\left.E\frac{\partial \bck(x)}{\partial k}\right|_x =
\frac{ix}{2}\Big(i\partial_x\com{\sigma_1}{\bck}-\left.\frac{\partial\vphi}{\partial x}\right|_k\com{\sigma_1}{\bck}-\halft k\acom{\sigma_1}{\bck}+m_1\sigma_3\bck+m_2\bck\sigma_3\Big)-\halft\com{\sigma_1}{\bck}-iE\left.\frac{\partial \varphi(x)}{\partial k}\right|_x \bck ,
\eeqa
where the derivatives of the phase $\vphi(x)$ are given by \eq{phiderx} and \eq{phiderk}. The result \eq{chider1}, however, actually holds independently of the definition of $\vphi(x)$ and therefore defines the transformation of the wave function for all $x$ (not only for $p^2>0$) once the phase is defined properly.

As we saw in the previous section, the wave function has the simple frame dependence \eq{krel} when the variable $s$ \eq{sdef} rather than $x$ is held fixed. With $s$ fixed,
\beq\label{sderphi}
\left.E\frac{\partial \bck(s)}{\partial k}\right|_s= \left.E\frac{\partial \bck(x)}{\partial k}\right|_x + E\left.\frac{\partial x(s)}{\partial k}\right|_s\frac{\partial\bck(x)}{\partial x} = \left.E\frac{\partial \bck(x)}{\partial k}\right|_x -\frac{kx}{E-V(x)}\frac{\partial\bck(x)}{\partial x} .
\eeq
According to \eq{boostop} the boosted state is an eigenstate of $P^\mu$. As a consequence, 
its wave function has the form \eq{chidef}, and it suffices to verify the $k$ independence of $\phi = \halft\sigma_1\acom{\sigma_1}{\Phi}$ suggested by \eq{bsephi}. Taking the anticommutator of \eq{sderphi} with $\sigma_1$ eliminates the terms with $\com{\sigma_1}{\bck}$ in \eq{chider1}, giving
\beq
\sigma_1\left.\frac{2E}{ix}\,\frac{\partial \phi(s)}{\partial k}\right|_s = 
-k\phi-m_1\sigma_3\halft\com{\sigma_1}{\bck}+m_2\halft\com{\sigma_1}{\bck}\sigma_3-
\frac{m_1^2-m_2^2}{p^2}(E-V)\sigma_1\phi+ki\partial_s\sigma_1\phi =0 ,
\eeq
where the vanishing of the \rhs\ may be verified using \eq{bsephi} and
\beq
\halft\com{\sigma_1}{\bck} = \frac{\sigma_1}{p^2}\big(m_1\slashed{p}\,\phi-m_2\phi\,\slashed{p} \big) .
\eeq
The frame independence of $\phi(s)$ establishes the $k$-dependence of the wave function $\Phi$ given by \eq{krel}.

%%%%%%%%%%%%%%%%%%%%%%%%%%%%%%%%%%%%%%%%%%%%%%%%%%%%%%%%
\section{Discussion} \label{discussion}
%%%%%%%%%%%%%%%%%%%%%%%%%%%%%%%%%%%%%%%%%%%%%%%%%%%%%%%%

We have studied the frame dependence of $e^-\mu^+$ QED bound states in $D=1+1$ dimensions. To our knowledge this is the first demonstration of an explicit and exact boost of a relativistic bound state defined at equal time of the constituents. Two essential conditions had to be fulfilled in order to make this possible. Firstly, we work in the Born approximation (no loops). The dynamics is then insensitive to the $\ieps$ prescription of the fermion propagators. With a prescription giving retarded propagation, where both positive and negative energy fermions move forward in time, it is possible to avoid ``spurious'' pair production due to $Z$-diagrams.\footnote{This is also how the single electron Dirac wave function can describe relativistic electrons bound in an external potential, even though those states have an indefinite number of $e^+e^-$ pair constituents. Such pairs may alternatively be avoided by quantizing at equal Light-Front time, $x^+=x^0+x^1$ \cite{Brodsky:1997de}.} True pair production is also suppressed due to the absence of loops. Secondly, the interaction had to be instantaneous in time (Coulombic), to avoid Fock states with any number of propagating photons. We ensured this by working in Coulomb gauge ($A^1=0$) and in $D=1+1$ dimensions. Gauss' law allows to express $A^0$ in terms of the fermion fields as in \eq{a0expr}, so that the dynamics can be expressed solely in terms of the fermion fields.

The fact that the Poincar\'e generators \eq{spacegen}, \eq{timegenqed} and \eq{boostgenqed} satisfy the Lie algebra \eq{plie} implies that the boosted state remains an eigenstate of the Hamiltonian with appropriately modified energy and CM momentum. The frame dependence \eq{krel} of the wave function, first noted in \cite{Hoyer:1986ei}, is remarkably simple yet enigmatic. The underlying reason for the emergence of the invariant length $s(x)$, defined by \eq{sdef}, is not clear from our derivation. The dependence on the kinetic energy $E-V(x) \sim i\partial^0-eA^0$ nevertheless seems natural in a gauge theory framework. 
Further studies are required to extend this simple formulation for boosting the wave function to cover the cases where the square of the ``momentum'' $p$ in \eq{pdef} is negative or zero, and hence to all values of the coordinate $x$.

Understanding the frame dependence of bound states is essential in studies of scattering amplitudes with bound states as external particles. The usefulness of the present approach will depend on its applicability to more physical systems, the relevance of the Born approximation and the possibility to calculate loop corrections. In Appendix \ref{qcdapp} we discuss the Poincar\'e generators of QCD$_2$. The non-abelian gauge invariance brings some new features, but the formulation of QCD bound states and their frame dependence is similar to the abelian case \cite{Hoyer:2009ep}. 

It is obviously more challenging to generalize the present approach to QCD bound states in $D=3+1$ dimensions. In order to avoid Fock states with any number of propagating transverse gluons the interaction should, in all frames, be dominated by instantaneous Coulomb exchange. The linear potential of QCD$_2$ was moreover essential for the closure of the Lie algebra in the non-local formulation involving only quark fields. It appears possible to apply the methods presented above also in $D=3+1$ dimensions by imposing a non-vanishing boundary condition on the solution of Gauss' law for $A^0$. This gives rise to a linear instantaneous potential in all frames, which is of \order{g} in the coupling and thus leading compared to the \order{g^2} transverse gluon exchange \cite{Hoyer:2009ep}. The frame dependence of the wave function turns out to be similar to the $D=1+1$ case when the quark positions are aligned with the CM momentum \cite{Hoyer:1986ei}. The other configurations of the wave function may then be solved numerically using the bound state equation.

\acknowledgements

Part of this work was done while the authors were visiting or employed by CP$^3$-Origins at the University of Southern Denmark. PH is grateful for the hospitality of CP$^3$-Origins as well as for a travel grant from the Magnus Ehrnrooth Foundation. The work of DDD was supported in part by the ExtreMe Matter Institute EMMI. The work of MJ was supported in part by the European grants FP7-REGPOT-2008-1: CreteHEPCosmo-228644 and PERG07-GA-2010-268246.

\appendix

%%%%%%%%%%%%%%%%%%%%%%%%%%%%%%%%%%%%%%%%%%%%%%%%%%%%%%%%
\section{Poincar\'e generators} \label{generators}
%%%%%%%%%%%%%%%%%%%%%%%%%%%%%%%%%%%%%%%%%%%%%%%%%%%%%%%%

The derivation of the generators of time translations $P^0$ and boosts $M^{01}$ is analogous to that of the space translation generator $P^1$ and is given below. We assume a single flavour $f$ for simplicity of notation.

\vspace{.5cm}

%%%%%%%%%%%%%%%%%%%%%%
\noindent{\it The generator $P^0$ of time translations}

\vspace{.3cm}

In a generic time translation the fermion field transforms as
\beq
 \psi(x^0,x^1) \to  \psi\big(x^0-\epsilon(x^0) dt,x^1\big) ,
\eeq
and the variation of the free fermion action \eq{qedact1} becomes
\beq
 \delta S_F =  -dt \int d^2x\ \psi^\dag(x)\gamma^0\big(i\slashed{\partial}-m\big)\epsilon(x^0) \partial_0 \psi(x) - dt \int d^2x\ \epsilon(x^0) \big(\partial_0 \psi^\dag(x)\big)\gamma^0\big(i\slashed{\partial}-m\big) \psi(x) \ .
\eeq
Integrating the last term partially over $x^0$ we find a contribution due to the dependence of $\epsilon$ on $x^0$,
\beq
 \delta S_F =  - dt \int d^2x\ \epsilon'(x^0)  \psi^\dag(x)\gamma^0\big(-i\gamma^1\partial_1 +m\big) \psi(x) \ .
\eeq
The variation of $S_V$ can be written as
\beqa
 \delta S_V &=&- \frac{dt e^2}{4}\int d^2x d^2y\ \delta(x^0-y^0) \epsilon(x^0) \frac{\partial}{\partial x^0}\left[\psi^\dag\psi(x)\right] |x^1-y^1|\psi^\dag\psi(y)\nn\\
&& - \frac{dt e^2}{4}\int d^2x d^2y\ \delta(x^0-y^0) \psi^\dag\psi(x) |x^1-y^1| \epsilon(y^0)  \frac{\partial}{\partial y^0} \left[\psi^\dag\psi(y)\right] \ .
\eeqa
A partial integration gives 
\beq
 \delta S_V  = \frac{dt e^2}{4}\int d^2x d^2y\ \epsilon'(x^0)  \delta(x^0-y^0) \psi^\dag\psi(x) |x^1-y^1| \psi^\dag\psi(y) \ .
\eeq
Collecting the results,
\beq
 \delta S = - dt \int dx^0\ \epsilon'(x^0) P^0 ,
\eeq
where $P^0$ is the Hamiltonian
\beq
 P^0 =  \int dx^1\ \psi^\dag(x)(-i\gamma^0\gamma^1\partial_1+m\gamma^0)\psi(x)-\frac{e^2}{4}\int dx^1 dy^1\  \psi^\dag\psi(x^0,x^1) |x^1-y^1| \psi^\dag\psi(x^0,y^1) \ .
\eeq
The covariance of the action and the conservation of $P^0$ follow 
as for space translations in the main text.

\vspace{.5cm}

%%%%%%%%%%%%%%%%%%%%%%
\noindent{\it The boost generator $M^{01}$}

\vspace{.3cm}

An infinitesimal boost in the $x^1$-direction which transforms the coordinates as 
\beq\label{coordboost}
x^0 \to x^0+d\xi x^1 , \hspace{2cm} x^1 \to x^1+d\xi x^0
\eeq
also generates an $A^1$ component of the gauge field: $(A^0,A^1=0) \to (A^0,d\xi A^0)$. In order to stay in the $A^1=0$ gauge we need to follow up with a gauge transformation 
\beq\label{gaugetrans}
\psi(x) \to \exp(-id\xi\,\theta)\psi(x) 
\eeq
with
\beq
\partial_1\theta(x) = eA^0(x)=-\frac{e^2}{2}\int d^2y\, \delta(x^0-y^0) |x^1-y^1| \psi^\dag\psi(y) ,
\eeq 
where $A^0$ was taken from \eq{a0expr}. This gives
\beq\label{gaugepar}
\theta(x) = -\frac{e^2}{4}\int d^2y\, \delta(x^0-y^0) (x^1-y^1)|x^1-y^1| \psi^\dag\psi(y) .
\eeq
Combined with the standard boost transformation we have then
\beq \label{combinf2}
 \psi(x^0,x^1) \to \left[1+\halft \epsilon(x^0)\gamma^0\gamma^1 d\xi - i \epsilon(x^0) \theta(x^0,x^1) d\xi \right]\, \psi\big(x^0-\epsilon(x^0) x^1 d\xi,x^1-\epsilon(x^0) x^0 d\xi\big) .
\eeq
We can decompose this into boost, spin, and gauge transformations, defined as
\beqa\label{qedboost2}
\psi(x^0,x^1) &{\buildrel{\mathrm{boost}}\over{\to}}& \psi\big(x^0-\epsilon(x^0) x^1 d\xi,x^1-\epsilon(x^0) x^0 d\xi\big) , \label{boost} \\ 
\psi(x^0,x^1) &{\buildrel{\mathrm{spin}}\over{\to}}& \left[1+\halft \epsilon(x^0)\gamma^0\gamma^1 d\xi\right]\psi(x^0,x^1) , \label{spin} \\ 
\psi(x^0,x^1) &{\buildrel{\mathrm{gauge}}\over{\to}}& \left[1 - i \epsilon(x^0) \theta(x^0,x^1) d\xi \right]\psi(x^0,x^1) . \label{gauge}
\eeqa
 
Since $S_F$ in \eq{qedact1} is explicitly Lorentz covariant we expect that its combined boost and spin transformation only involves terms with $\epsilon'(x^0)$. A
straightforward calculation gives
\beqa
 \delta_\mathrm{boost} S_F +\delta_\mathrm{spin} S_F &=& \halft i d\xi \int d^2x \ \eps'(x^0) \psi^\dag(x) \gamma^0\gamma^1 \psi(x) \nn\\ 
&&-d\xi \int d^2x \ \eps'(x^0)  \psi^\dag(x) \big(ix^0\partial_1 - i x^1 \gamma^0\gamma^1 \partial_1 + x^1 \gamma^0 m \big)\psi(x) \ . 
\eeqa
The variation under the gauge transformation \eq{gauge} is
\beqa\label{sfgaugedep}
 \delta_\mathrm{gauge} S_F &=& d\xi \int d^2x\ \psi^\dag(x)\gamma^0\big[\gamma^\mu \partial_\mu \epsilon(x^0) \theta(x)\big]\psi(x) \\
&=& -\frac{d\xi e^2}{2} \int d^2xd^2y\ \epsilon(x^0) \delta(x^0-y^0) \psi^\dag(x) \gamma^0\gamma^1 \psi(x)|x^1-y^1| \psi^\dag\psi(y) \nn\\
&& - \frac{d\xi e^2}{4}\int d^2x d^2y\ \epsilon(x^0)\delta(x^0-y^0) (x^1-y^1)|x^1-y^1| \psi^\dag \psi(x)\frac{\partial}{\partial y^0} \psi^\dag\psi(y) , \nn
\eeqa
where the first (second) term arises from the spatial (time) derivative of $\theta(x)$ in \eq{gaugepar}. The contribution involving $\eps'(x^0)$ vanishes due to the antisymmetry of the integrand under $x \leftrightarrow y$.\footnote{The fermion fields at $x$ 
anticommute with those at $y$ due to the factor $\delta(x^0-y^0)|x^1-y^1|$.} 

The fields in the potential term $S_V$ of \eq{qedact1} appear in the gauge invariant combination $\psi^\dag(x) \psi(x)$. Thus 
\beq\label{dsvzero}
\delta_\mathrm{gauge} S_V=0 .
\eeq
The spin transformation becomes
\beq\label{spincont2}
 \delta_\mathrm{spin} S_V = \frac{d\xi e^2}{2} \int d^2x d^2 y\ \epsilon(x^0) \delta(x^0-y^0) \psi^\dag \gamma^0\gamma^1 \psi(x)|x^1-y^1| \psi^\dag\psi(y)
\eeq
after using the symmetry of the integration measure under $x \leftrightarrow y$. In the boost transformation we notice that since initially $x^0=y^0$ in $S_V$, the shift of the space coordinate is the same for all fields, and can be absorbed into a shift $d \xi x^0$ of the integration variables $x^1, y^1$ similarly as in the treatment of the space translations. 
The remaining contribution from the transformation of the time coordinates can be expressed as
\beqa\label{boostcont3}
 \delta_\mathrm{boost} S_V &=& - \frac{d \xi e^2}{4} \int d^2x d^2 y\ \delta(x^0-y^0)|x^1-y^1| x^1 \epsilon(x^0) \frac{\partial}{\partial x^0} \psi^\dag \psi(x)\psi^\dag \psi(y) \\
&& -\frac{d \xi e^2}{4} \int d^2 x d^2 y\ \delta(x^0-y^0)|x^1-y^1| y^1 \psi^\dag \psi(x) \epsilon(y^0)  \frac{\partial}{\partial y^0}\psi^\dag \psi(y) \\
 &=&  \frac{d \xi e^2}{4}  \int d^2x d^2 y\ \epsilon(x^0) \delta(x^0-y^0)(x^1-y^1)|x^1-y^1|  \psi^\dag \psi(x)  \frac{\partial}{\partial y^0}\psi^\dag \psi(y) \nn\\
&& +\frac{d\xi e^2}{4} \int d^2 x d^2 y\ \eps'(x^0) x^1 \psi^\dag\psi(x) \delta(x^0-y^0) |x^1-y^1| \psi^\dag\psi(y) ,
\label{boostcont4}
\eeqa
where the latter expression was obtained by partial integration. Adding up the various contributions, only terms involving the derivative of $\epsilon(x^0)$ survive:
\beq \label{finalvar2}
 \delta S = d \xi \int dx^0\ \eps'(x^0)  M^{01} ,
\eeq
where, using again the $x \leftrightarrow y$ symmetry of the integration measure in the potential term,
\beqa
 M^{01} &=& x^0\int dx^1\ \psi^\dag(x)(-i\partial_1)\psi(x) +  \int dx^1\ \psi^\dag(x)\Big[ x^1( i \gamma^0\gamma^1 \partial_1 - \gamma^0 m) + \frac{i}{2} \gamma^0\gamma^1 \Big] \psi(x) \nn\\
&&+\frac{e^2}{8}\int dx^1 dy^1\  \psi^\dag\psi(x^0,x^1) (x^1+y^1)|x^1-y^1| \psi^\dag\psi(x^0,y^1) \ .
\eeqa
In terms of the momentum densities
\beqa
{\mathcal P}^0 &=& \bar\psi\big(-\halft i\gamma^1\lrder_1+m\big)\psi 
-\frac{e^2}{4}\int dy^1\  \psi^\dag\psi(x^0,x^1) |x^1-y^1| \psi^\dag\psi(x^0,y^1) , \nn \\
{\mathcal P}^1 &=&  \bar\psi \big(-\halft i \gamma^0 \lrder_1 \big)\psi ,
\eeqa
the boost density has the expected form,
\beq
{\mathcal M}^{01} = x^0 {\mathcal P}^1 - x^1 {\mathcal P}^0 \ .
\eeq

%%%%%%%%%%%%%%%%%%%%%%%%%%%%%%%%%%%%%%%%%%%%%%%%%%%%%%%%%%%%%%%%
\section{Poincar\'e algebra of QCD$_2$} \label{qcdapp}
%%%%%%%%%%%%%%%%%%%%%%%%%%%%%%%%%%%%%%%%%%%%%%%%%%%%%%%%%%%%%%%%

The derivation of the QED$_2$ generators in Appendix \ref{generators} can be carried out similarly for QCD$_2$. In Coulomb gauge the solution of Gauss' law (without a constant background field, and for a single flavor) gives
\beq\label{a0aexpr}
A_a^0(x) = -\frac{g}{2}\sum_{A,B} \int d^2y \, \delta(x^0-y^0) |x^1-y^1| \psi_A^\dag(y) T_a^{AB}\psi_B(y) .
\eeq
Substituting this expression in the QCD$_2$ action gives (suppressing the quark color indices)
\beq\label{qcdact1}
S =\int d^2x\, \psi^\dag(x)\gamma^0\big(i\slashed{\partial}-m\big)\psi(x) +
\frac{g^2}{4}\sum_{a}\int d^2x\, d^2y\, \delta(x^0-y^0) \psi^\dag(x) T_a\psi(x) |x^1-y^1| \psi^\dag(y) T_a\psi(y) .
\eeq
The free parts of the QCD generators ($P^1,\ P^0_F,\ M^{01}_F$) are the same as the corresponding QED generators given in \eq{spacegen}, \eq{timegenqed} and \eq{boostgenqed}, when a sum over the quark color indices is understood. The QCD interaction terms are
\beqa
P^0_V &=& -\frac{g^2}{4}\sum_{a}\int dx^1 dy^1\  \psi^\dag T_a \psi(x^0,x^1) |x^1-y^1| \psi^\dag T_a \psi(x^0,y^1) \label{timegenqcd} , \\[2mm]
M^{01}_V &=& \frac{g^2}{8}\sum_{a}\int dx^1 dy^1\  \psi^\dag T_a \psi(x^0,x^1) (x^1+y^1)|x^1-y^1| \psi^\dag T_a \psi(x^0,y^1) . \label{boostgenqcd}
\eeqa
The boost operator $M^{01}$ is again a combination of a boost and a gauge transformation \eq{gaugetrans} which ensures that $A_a^1=0$ after the boost. The gauge parameter corresponding to \eq{gaugepar} is
\beq\label{gaugeparqcd}
\theta_a(x) = -\frac{g^2}{4}\int d^2y\, \delta(x^0-y^0) (x^1-y^1)|x^1-y^1| \psi^\dag(y) T_a\psi(y) .
\eeq

The Poincar\'e Lie algebra \eq{plie} works out as in QED, with one exception. Due to the color generators $T_a$ in $P^0_V$ and $M^{01}_V$ these two terms do not commute in QCD. Hence the Lie algebra does not close, as already observed in \cite{Bars:1977ud}. In the shorthand 
\beq\label{phidefqcd}
\phi_a(x) \equiv \psi^\dag(x)T^a\psi(x)
\eeq
we have
\beq\label{phicomm}
\com{\phi_a(x)}{\phi_b(y)} = i f_{abc}\phi_c(x)\delta(x-y) ,
\eeq
where a sum over repeated color indices is understood. Since all operators are evaluated at a common time we here and in the following denote $x \equiv x^1,\ etc.$ We then find
\beqa\label{plieqcd}
\com{P^0}{M^{01}}-iP^1=\com{P^0_V}{M^{01}_V} &=& -\frac{g^4}{32}\int dx dy du dv\com{\phi_a(x)\phi_a(y)}{\phi_b(u)\phi_b(v)}(u+v)|u-v||x-y| \nn\\[2mm]
&=& i\frac{g^4}{48}f_{abc}\int dx dy du\, \phi_a(x)\phi_b(y)\phi_c(u)(u-x)(u-y)(x-y) .
\eeqa
In obtaining the final expression we made repeated use of the fact that commutators like \eq{phicomm} vanish when multiplied by $x-y$.

The non-closure of the Lie algebra is related to the gauge transformation embedded in $M^{01}$, which implies that operators occurring before the boost are in a different gauge compared to those occurring after the boost. If we consider the commutation relation of the {\em group elements} rather than the generators we have
\beq\label{groupcom}
\exp(-idt\,P^0_{new})\exp(id\xi\,M^{01})-\exp(id\xi\,M^{01})\exp(-idt\,P^0_{old})=dt\,d\xi\com{P^0}{M^{01}}-idt(P^0_{new}-P^0_{old})= idtd\xi\, P^1 .
\eeq 
Here $new$ and $old$ refer to the operators after and before the gauge transformation\footnote{The gauge dependence refers only to the interaction terms $P^0_V$ and $M^{01}_V$. The free parts of the generators satisfy the Lie algebra as in QED.}, \ie,
\beqa
\psi_{new} &=& (1-id\xi\theta_a T_a)\psi_{old} , \nn\\[2mm]
\phi_a^{new}(x)-\phi_a^{old}(x)&=& d\xi\, f_{abc} \psi^\dag(x) \theta_b(x)T_c\psi(x) ,
\eeqa
with $\theta_b$ given by \eq{gaugeparqcd}. The gauge dependence of the infinitesimal time translation in \eq{groupcom} is then
\beqa\label{gaugecorr}
-idt\big(P^0_{V,new}-P^0_{V,old}\big) &=& idt d\xi \frac{g^2}{4}f_{abc}\int dx dy\big\{ \psi^\dag(x) \theta_b(x)T^c\psi(x)\,\phi_a(y)+\phi_a(x)\,\psi^\dag(y) \theta_b(y)T^c\psi(y)\big\}|x-y|\nn\\
&=& -idt d\xi \frac{g^4}{48}f_{abc}\int dx dy du\, \phi_a(x)\phi_b(y)\phi_c(u)(u-x)(u-y)(x-y)
\eeqa 
According to \eq{plieqcd} this term cancels against  
$dt\,d\xi\com{P^0_V}{M^{01}_V}$ in the commutator \eq{groupcom} of the group elements, giving the pure translation already indicated in \eq{groupcom}.

The generalization of the QED bound states considered in this paper to those of QCD is straightforward, and is unaffected by the contribution \eq{plieqcd}. In analogy to \eq{ketk} for QED we define a color-singlet meson state by
\beq \label{ketkq}
 \ket{E,k}\equiv \sum_{A,B}\int dx_1 dx_2\, \exp\big[\halft ik(x_1+x_2)+i\vphi(x_1-x_2)\big]\bar\psi_{1A}(0,x_1)\delta_{AB}\bck(x_1-x_2)\psi_{2B}(0,x_2)\ket{0}_R . 
\eeq
Requiring this to be an eigenstate of the QCD Hamiltonian gives the bound state condition \eq{bse} on the wave function $\bck(x)$. The potential generated by $P_V^0$ in \eq{timegenqcd},
\beq
V(x)=\halft g^2 C_F |x| ,
\eeq
includes the expected color coefficient $C_F=(N^2-1)/2N$. The contribution \eq{plieqcd} to $\com{P^0}{M^{01}}-iP^1$ annihilates on the bound state \eq{ketkq}, ensuring that the frame dependence of the wave function is the same as in QED. This is consistent with the observation of \cite{Bars:1977ud} that the mismatch of the Lie algebra in \eq{plieqcd} is proportional to the charge operator.

\end{document}